\documentclass[10pt,twocolumn,letterpaper]{article}

\usepackage{soul}
\usepackage[pagenumbers]{cvpr}

\definecolor{cvprblue}{rgb}{0.21,0.49,0.74}
\usepackage[pagebackref,breaklinks,colorlinks,allcolors=cvprblue]{hyperref}

\def\paperID{AI4Space-28}
\def\confName{CVPR}
\def\confYear{2026}

\title{Thermal Anomaly Detection using Physics Aware Neuromorphic Networks: Comparison between Raw and L1C Sentinel-2 Data}

\author{Stephen Smith$^1$ \\
\and
Cormac Purcell$^{1,2,3}$\\
\and
Gabriele Meoni$^{4,5}$\\
\and
Roberto Del Prete$^4$\\
\and
Zdenka Kuncic$^{1,6}$\\
\and
$^1$ School of Physics, University of Sydney, Sydney, Australia\\
$^2$ BolgianTen Ltd., Liverpool, United Kingdom\\
$^3$ School of Computer Science, University of New South Wales, Sydney, Australia\\
$^4 \Phi$-lab, European Space Research Institute (ESRIN), European Space Agency (ESA), Frascati, Italy\\
$^5$ Advanced Concepts and Studies Office, European Space Research Institute (ESRIN), \\
European Space Agency (ESA), Frascati, Italy\\
$^6$ Emergentia Pty. Ltd., Sydney, Australia\\
}

\begin{document}
\maketitle
\begin{abstract}
Damage caused by bushfires and volcanic eruptions escalates rapidly when detection is delayed, making fast and reliable early warning capabilities essential. Recent Earth Observation (EO) approaches have shown that thermal anomaly detection can be performed directly on decompressed Level--0 (L0) sensor data, avoiding computationally expensive preprocessing chains. However, direct exploitation of raw data remains challenging due to domain shift, sensor drift, radiometric inconsistencies, and the scarcity of labelled training samples.
To address these challenges, this work proposes a Physics-Aware Neuromorphic Network (PANN) framework for onboard thermal anomaly detection. The proposed lightweight architecture, inspired by physical neural network principles and neuromorphic computing paradigms, is evaluated using two Sentinel-2 datasets: decompressed L0 with additional metadata (i.e. raw) and Level--1C (L1C). The PANN achieves a Matthews Correlation Coefficient (MCC) of $0.809$ on raw measurements, compared to $0.875$ when using ground-processed L1C products. The mean processing latency per L0 granule is $2.44 \pm 0.09~\mathrm{s}$, which is below the Sentinel-2 acquisition time of $3.6~\mathrm{s}$, demonstrating the feasibility of real--time, onboard processing. Furthermore, the projected execution time for the corresponding neuromorphic hardware instantiation is substantially lower at $0.1290 \pm 0.0002~\mathrm{s}$.
Memory usage, including all necessary programs and packages, remains within realistic onboard constraints, with requirements of $0.673 \pm 0.007~\mathrm{Gb}$ for the software PANN and $0.393 \pm 0.004~\mathrm{Gb}$ for the estimated hardware realisation. Overall, these results indicate that PANN offers a promising pathway toward low-latency and resource-efficient onboard EO processing for thermal event detection.
\end{abstract}    
\section{Introduction}
\label{sec:intro}

Thermal hotspots or anomalies, such as bushfires and volcanic activity, can cause severe environmental and societal impacts\cite{bowman2017human,bowman2020vegetation}, including adversely impacting air and water quality, ecosystems and human welfare \cite{paul2022wildfire, jaffe2020wildfire, durant2010atmospheric}. 
Thus, it is vital that thermal hotspots and anomalies are detected and reported to emergency services before they become out of control \cite{toan2019deep}, especially as fire seasons become longer, with more intense events \cite{flannigan2013global}. While Earth observation (EO) solutions can provide the necessary coverage, for the information to be most effective, latency must be minimised between data acquisition (with limited revisit times), thermal hotspot detection and arrival of information to decision makers and first responders. An approach to this problem is to move the image processing and thermal event detection to be performed onboard the EO satellite.

In missions such as Sentinel-2 \cite{drusch2012sentinel} and LandSat-8 \cite{wulder2019current}, multispectral EO images are captured with pushbroom based sensors and require significant processing to enhance image quality. Here, we focus on Sentinel-2 data, but the fundamental processing operations are applied to other data processing chains \cite{gatti2015sentinel}, even though implementation of such processing chains and the corresponding product nomenclature vary between missions.
The Sentinel-2 satellites acquire EO images using a multispectral imager \cite{drusch2012sentinel}. The sensor has a total of 12 detectors that, in a single acquisition, each capture a granule for each spectral band in 3.6\,s in a pushbroom motion \cite{gatti2015sentinel}.
These bands are not aligned due to the pushbroom nature of the images and along track parallax error. The images are equalised and compressed before being downlinked to earth, and are defined as the decompressed level-0 (L0) data with additional metadata, often referred as the ``raw'' data \cite{meoni2024unlocking}. 
Once on the ground, the raw data undergoes several processing steps, e.g. spatial coregistration of the spectral bands, radiometric correction, invert onboard equalisation, adding the radiometric corrected geometric model, orthorectification and calculation of top of atmosphere reflectances, among others \cite{meoni2024unlocking}. At this stage the Sentinel-2 data is defined as level-1C (L1C) data. 
This is the lowest level data that is routinely made publicly available from Copernicus Data Space Ecosystem and ESA projects, with lower level products such as Level-1B released to expert users only from 2025.

Typically, studies that have involved processing EO images, particularly for detecting natural disasters, have used the L1C data or higher level processed data as this is what is publicly available. With the growing need for onboard processing, recent studies have begun to shift attention to the raw L0 data \cite{meoni2025e2e,del2025enhancing,traba2026towards}, to avoid the need to perform the computationally intensive processing required to transform the data from raw to L1C.
As such, AI models that are to be deployed onboard EO satellites need to be able to process the raw data, while operating under tight resource constraints present on satellites. 

While some models address the need to handle raw data, all models require training on the ground before being deployed on satellites to perform inference. The first challenge is that machine learning models need labelled training data and often do not generalise well to new data from a different sensor, resulting in a significant decrease in performance 
\cite{koh2021wilds}, therefore requiring data and labels to be gathered on an ongoing basis for new sensors and missions. 
Using simulations is an option, but is often time consuming and expensive \cite{chintalapati2025opportunities}. Over the duration of a mission, sensors can drift or the domain problem can shift, requiring models to be retrained with new data to maintain performance \cite{mateo2023orbit}. 

This study investigates a new approach to processing raw data using a Physics Aware Neuromorphic Network (PANN), an AI model based on physical neural networks \cite{KuncicNakayama2021,zhu2021information, vahlBraininspired2024, caravelli2025self}. These are effectively hardware neural networks with a brain-inspired, recurrent architecture comprised of self-assembled nanowires that form nano-electronic circuit elements known as memristors (nonlinear resistors with memory) \cite{Caravelli2018curious}. The physical nanowire network substrate embeds memristive weights that continuously update their states in response to varying input signals according to a set of physical laws (a memristor equation of state and Kirchhoff's circuit laws). As such, the PANN actively extracts dynamical features from each new input without training any network weights. This opens up the possibility of energy-efficient continual learning on-orbit. In contrast, this is not possible with neuromorphic approaches that typically implement spike-based point neuron models, which require significant data preprocessing and model training \cite{lunghi2025SNN}. 
The PANN model was previously applied to an EO natural disaster change detection task \cite{smith2025training}, where the PANN processed a sequence of images from the Sentinel-2 satellites using L1C data. This study extends that work by applying the PANN model to thermal event detection using newly available Sentinel-2 raw data and comparing model performance to corresponding L1C data.

The remainder of this paper is organised as follows: \cref{sec:method} details the datasets used and presents the PANN model, along with the entire pipeline. \Cref{sec:resultsAndDiscussion} compares the performance of the PANN model on the raw and L1C datasets used, as well as showing the time and memory requirements of the model and estimates for the hardware equivalent. \Cref{sec:conclusion} presents a summary the main findings of this study and suggestions for future work.
\section{Methodology}
\label{sec:method}

This study used a modified THRawS dataset \cite{meoni2024unlocking} containing Sentinel-2 images of thermal hotspots with two levels of processed data, as detailed in \cref{sec:dataset}. Details of the PANN model and parameters used are given in \cref{sec:modelArch}, with the full model pipeline given in \cref{sec:modelPipeline}. \Cref{sec:metrics} describes the method and hardware used to measure the inference time and memory requirement of the model.

\subsection{Dataset}
\label{sec:dataset}

The THRawS dataset used in this study contains images of thermal hotspots caused by bushfires and volcanoes, from around the world. In addition, it also contains images of inactive volcanoes to ensure models are learning to detect thermal hotspots, not volcano calderas. For this study, we only used two of the spectral bands, B8A in the near infrared and B12 in the short wave infrared. For each granule location in the dataset there is a raw image and a processed L1C image. Each raw granule had a coarse coregistration \cite{meoni2024unlocking, meoni2025e2e} applied and adjacent granules were used to fill in the missing data, due to the coarse coregistration of the spectral bands. When no adjacent granule was available, the granule was cropped. This allowed as much of the dataset as possible to be kept, while ensuring that all inputs into the model had both bands present. To keep the raw and L1C datasets the same, the corresponding L1C granules were also cropped and the SuperGlue neural network \cite{sarlin2020superglue} was used to align the L1C granules with the raw granules.

Granules, which are $1152\times 1296$ pixels, were tiled into smaller $128\times 128$ pixel patches with no overlap. The patches were automatically labelled using the warm thermal hotspot algorithm by  Massimetti et al. \cite{massimetti2020volcanic} on the L1C data. Patches were considered to be an ``event'', if a hotspot contained at least 9 pixels, to be consistent with \cite{meoni2024unlocking}. To complete the labelling process, all patches were visually inspected, any patches that contained errors were removed. The final dataset had a total of 21,450 patches of which 442 (2\%) were ``events'' and 21,008 (98\%) were ``non-events'', for both the raw and L1C datasets.
 
To account for the large class imbalance in the dataset, the Matthew’s Correlation Coefficient (MCC) was used to evaluate the performance of the model and is defined as:

{\small
\begin{equation}
  \mbox{MCC} = \frac{T_N \cdot T_P - F_N \cdot F_P}{\sqrt{(T_P+F_P)(T_P+F_N)(T_N+F_N)(T_N+F_P)}}
  \label{eq:mcc}
\end{equation}
}
where $T_P$, $T_N$, $F_P$ and $F_N$ are the total number of True Positives, True Negatives, False Positives and False Negatives, respectively. The precision and recall of the model's performance were also calculated as:

\begin{equation}
    \text{precision} = \frac{T_P}{T_P + F_P} \qquad , 
\end{equation}

and 

\begin{equation}
    \text{recall} = \frac{T_P}{T_P + F_N} \qquad .
\end{equation}

\subsection{PANN Model Architecture}
\label{sec:modelArch}

PANN is based on a physically--motivated model of a nanowire network hardware device comprised of resistive switching memory (memristive) electrical junctions~\cite{Stieg2011,kuncicNeuromorphic2020a}. 
This is done by first modelling the nanowire network structure to create the topology of the PANN. To create the PANN network used in this study, 1520 nanowires were distributed over a $250 \times 250~\mu\mathrm{m}$ 2D plane with nanowire centres sampled from a generalised normal distribution with a beta value of 5 and with nanowire orientations sampled on $[0,\pi]$. Nanowire lengths were sampled from a Gaussian distribution with an average of $30~\mu\mathrm{m}$ and a standard deviation of $6~\mu\mathrm{m}$. As memristive nanowire networks are nano-electronic systems, input and output signals are delivered and read out via electrodes. Thus, a device was modelled by overlaying an evenly spaced $16\times 16$ electrode grid centred over the 2D plane. Each electrode had a $8~\mu\mathrm{m}$ diameter and a pitch of $8~\mu\mathrm{m}$ between electrodes. A quarter of the electrodes were selected in an evenly spaced grid pattern as the input electrodes, thereby forming an input layer. The remaining electrodes were used as readout electrodes forming an output layer. Where the nanowires overlap with other nanowires or electrodes, electrical junctions are formed, resulting in a total of 12,736 memristive junctions for the PANN network used in this study. 

\begin{figure}
    \centering
    \includegraphics[width=0.95\linewidth]{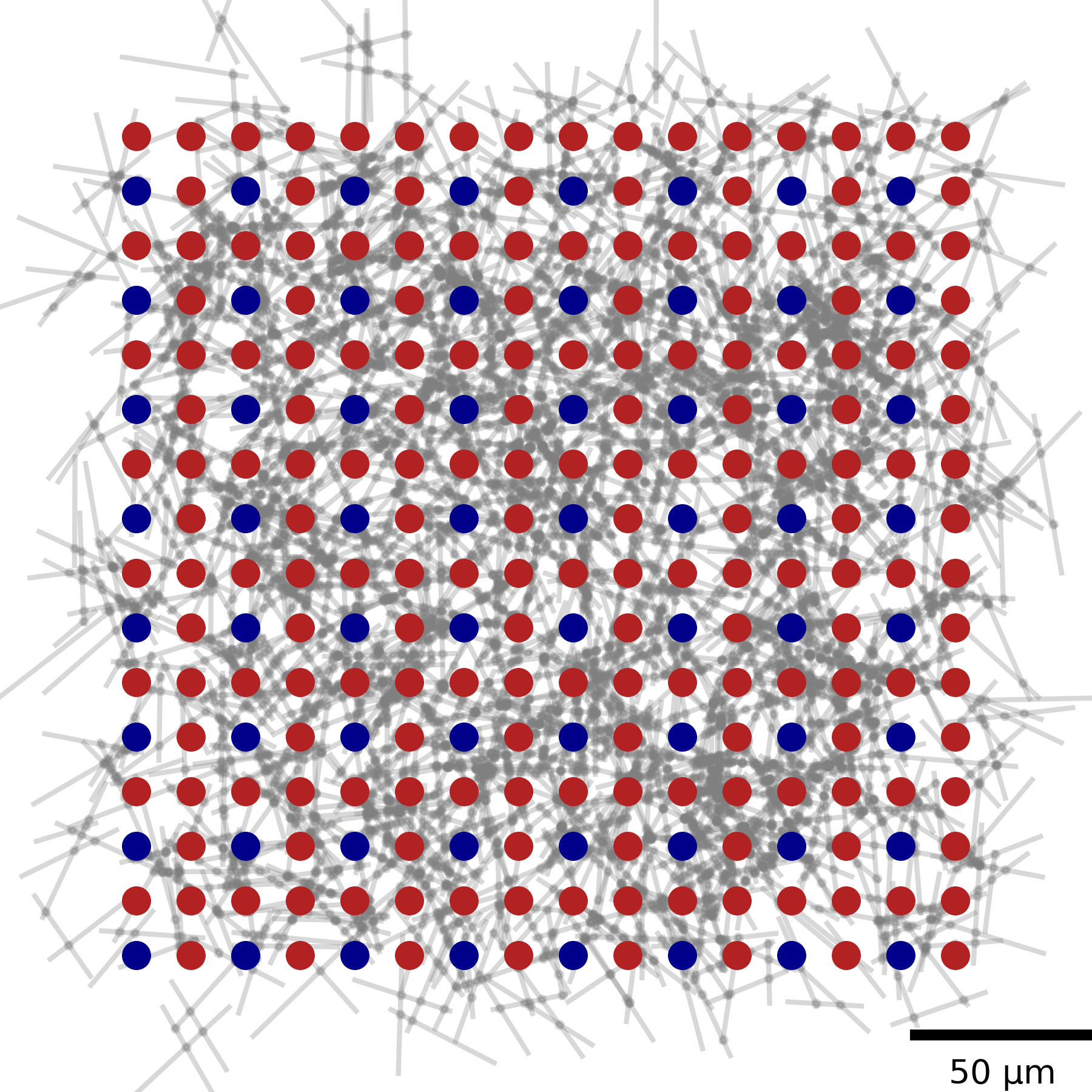}
    \caption{Visualisation of the simulated nanowire network device model used in PANN, showing the input electrodes (large blue circles), readout electrodes (large red circles) and the underlying nanowire network underneath with the nanowires (grey lines) and nanowire--nanowire junctions (small grey dots). Scale bar is $50~\mu\mathrm{m}$.}
    \label{fig:network}
\end{figure}

\Cref{fig:network} depicts the simulated nanowire network used for the PANN model. The nanowires and nanowire--nanowire junctions are displayed as grey lines and small grey dots, respectively. The electrodes are shown as the large circles, with the input electrodes in blue and readout electrodes in red. 

Inputs are treated as input voltage signals and Kirchhoff's conservation laws are solved at each time step to calculate the voltage distribution across the network. To do this, the network is converted to its graphical representation, with nanowires and memristors represented as nodes and edges, respectively. As the internal states of the memristive junctions evolve in time, the conductance of each junction can be considered as the weight of an edge in a neural network. 
As such, the weights in the PANN model dynamically evolve with each input and are not tuned using backpropagation or other weight-training algorithms. Thus, the network extracts features without requiring training before being used for inference.
The junction conductances evolve over time according to a memristor equation of state~\cite{kuncicNeuromorphic2020a}:

\begin{equation}
    \frac{d\lambda}{dt} = \begin{cases}
        (|V(t)| -V_{\text{set}})\text{sgn}[V(t)] , & |V(t)| > V_{\text{set}} \\
        0 , & V_{\text{reset}} < |V(t)| < V_{\text{set}} \\
        b(|V(t)| - V_{\text{reset}})\text{sgn}[\lambda(t)] , & |V(t)| < V_{\text{reset}} \\
        0 , & |\lambda| \ge \lambda_{\text{max}} \\
    \end{cases}
\label{eqn:filamentState}
\end{equation}

with all junctions set to an initial state $\lambda(t=0) = 0$.
In this study, the following parameters were used: $V_{\text{reset}} = 5\times10^{-3}~\mathrm{V}$, $V_{\text{set}} = 1\times10^{-2}~\mathrm{V}$, $\lambda_{\text{max}} = 1.5\times10^{-2}~\mathrm{V~s}$ and $b = 10$, which is a memory decay parameter.

The physics within the PANN model is therefore based on the physical nanowire network substrate, which nonlinearly transforms and integrates information by electrical switching processes that are history dependent, effectively extracting compact features from the data.
This is fundamentally different from physics informed neural networks, where the physics is specific to the domain problem and is typically introduced into an objective function used for training.

\subsection{Model Pipeline}
\label{sec:modelPipeline}

\begin{figure*}
  \centering
  \includegraphics[width=0.95\linewidth]{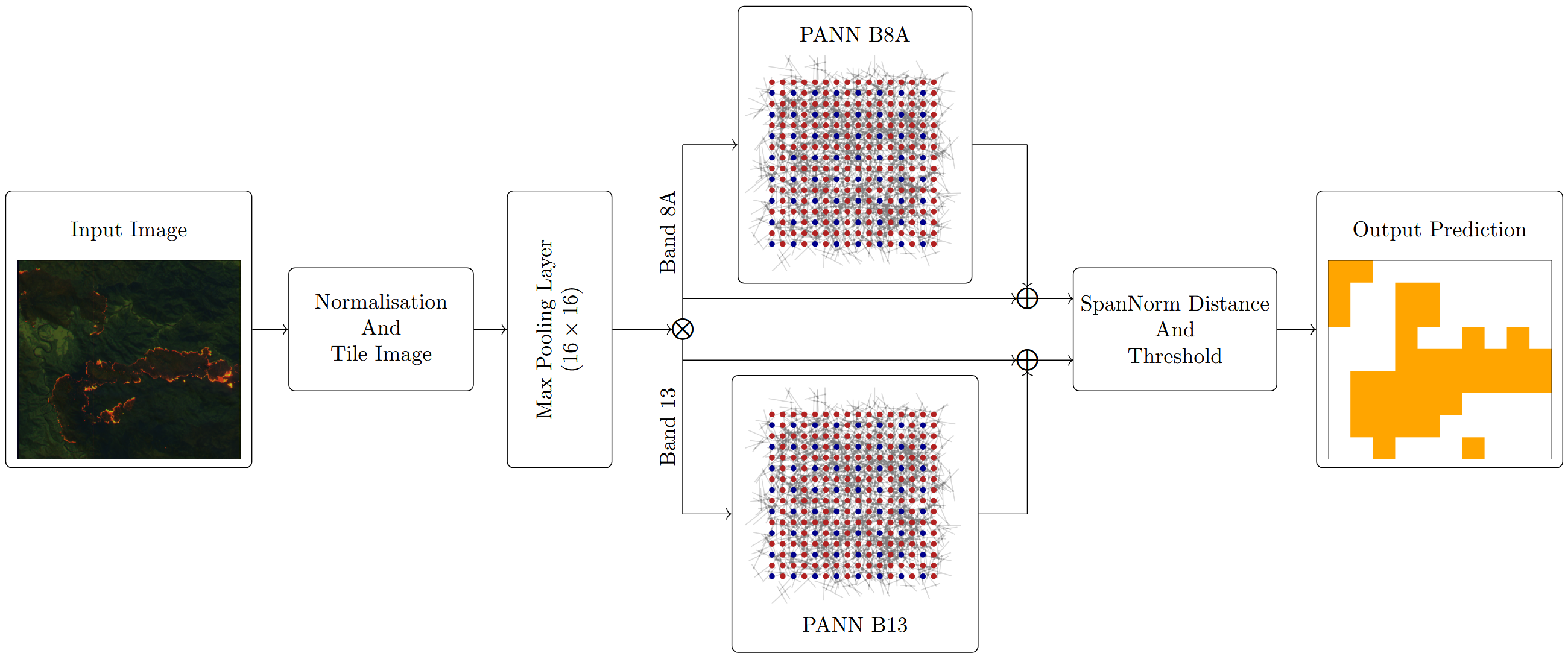}
  \caption{Diagram of the model pipeline. Starting with the input image (left), the image is fist normalised and tiled into smaller $128\times 128$ pixel patches, followed by a $16\times 16$ max pooling layer. The two spectral bands, B8A and B13, are split as denoted by $\bigotimes$ and fed into identical PANNs. Output feature vectors are concatenated with a skip connection, denoted by $\bigoplus$. Finally, the spanning normal distances between the two feature vectors is calculated and a threshold value is applied to classify each tile and produce a predicted output map (right).}
  \label{fig:modelSetup}
\end{figure*}

\Cref{fig:modelSetup} shows the full model pipeline. First, each band in the data is normalised according to \cref{eqn:normalisation}, 
\begin{equation}
    x = 1.2 \left[ \frac{x}{\mbox{max}(x) } \right] -0.4 \qquad ,
\label{eqn:normalisation}
\end{equation}
and scaled to the interval [-0.4,0.8].

The maximum value used for the different bands was based on values used in other works that used Sentinel-2 data \cite{ruuvzivcka2022ravaen, smith2025training}. In the raw data a max value of 3000 was used for both bands and values of 4 and 2 for bands B8A and B12 in the L1C data, respectively. Next, granules were tiled into smaller $128\times 128$ pixel patches with no overlap and a max pooling layer was applied with a pool size of $16\times 16$ and a stride of 16. 
The two bands were then split and fed into separate PANNs, each of which were interfaced via a $16\times 16$ electrode grid; with the input electrodes functioning as the input layer and the rest of the electrodes used as the readout layer.
As such, a feature vector is readout from the PANNs for each tile fed into the PANNs. A skip connection is used, to combine the inputs into the PANN and with the output feature vector, to create the final feature vector. The PANNs used to process each band were identical. Finally, the spanning normal distance, given by \cref{eq:spanNorm},
\begin{equation}
  \text{SpanNorm}(\vec{x},\vec{y}) = \text{max}(\vec{x}-\vec{y}) - \text{min}(\vec{x}-\vec{y})
  \label{eq:spanNorm}
\end{equation}
between the feature vectors of the two bands is calculated.
If the distance is above a predefined threshold, 1.68 and 0.92 for the raw and L1C datasets, respectively, that tile is classified as an ``event'' tile, otherwise it is classified as a ``non-event'' tile.

\subsection{Compute Metrics}
\label{sec:metrics}

All processing was performed on an Intel$^\circledR$ Xeon$^\circledR$ Platinum 8268 Processor using only 16 of the available 24 CPU cores.
The total time and maximum memory required for the PANN model, including all necessary programs and packages, to process each granule was calculated and then averaged across 5 granules. In addition, a time and memory usage estimate for the hardware based nanowire network was also calculated by performing all the steps shown in \cref{fig:modelSetup} excluding the PANN simulation, which was instead replaced with randomly generated numbers to replicate the output of the nanowire network. The hardware execution time, that is the time period being simulated by the PANN model, was also added on to give the estimated time. As with the PANN model, the estimated time and memory usage for the hardware nanowire network was averaged over 5 granules.
\section{Results and Discussion}
\label{sec:resultsAndDiscussion}

\subsection{Model Evaluation}

The performance of the model measured using the MCC, precision and recall for both the raw and L1C datasets is reported in \Cref{tab:results}. Since the PANN does not need to be trained, the entire dataset was used for performance testing. For MCC, the L1C data performed better with a value of 0.875, however, the raw data is not far behind at 0.809. Thus, the extra cleaning performed on the L1C data has helped to remove noise and increase performance. The relatively small difference between the L1C and raw MCC scores (0.066) shows that the PANN model is capable of extracting meaningful features from the noisier raw data for thermal hotspot detection.

\begin{table}
  \caption{MCC, precision and recall results for the PANN model for both raw and L1C datasets.}
  \label{tab:results}
  \centering
  \begin{tabular}{@{}lcc@{}}
    \toprule
     & Raw Data & L1C Data \\
    \midrule
    MCC & 0.809 & 0.875 \\
    Precision & 0.763 & 0.887\\
    Recall & 0.867 & 0.869\\
    \bottomrule
  \end{tabular}
\end{table}

\begin{figure}
    \centering
    \begin{subfigure}{0.95\linewidth}
        \includegraphics[width=\linewidth]{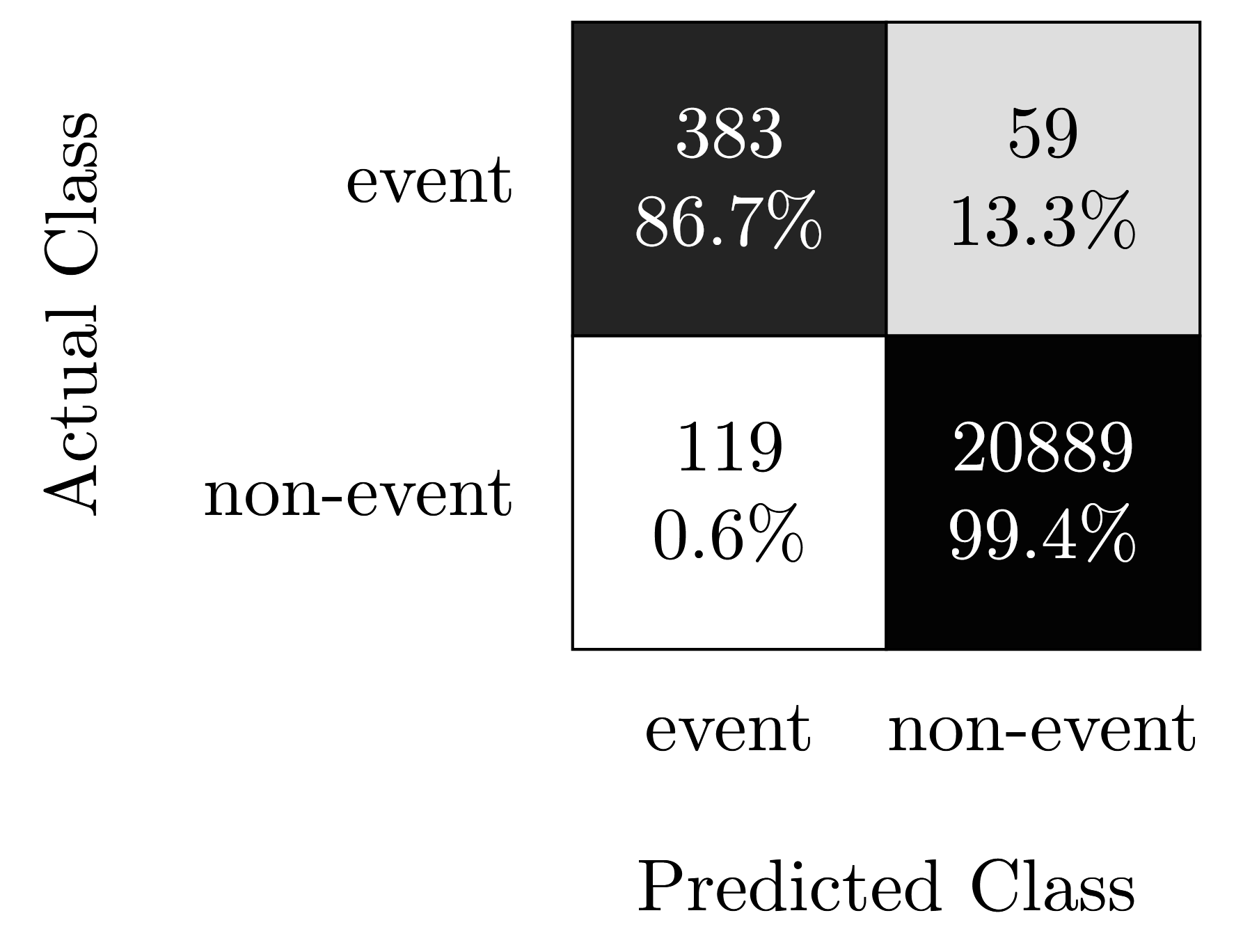}
    \caption{Confusion matrix for the raw data.}
    \label{fig:L0conMatrix}
    \end{subfigure}
    \begin{subfigure}{0.95\linewidth}
        \includegraphics[width=\linewidth]{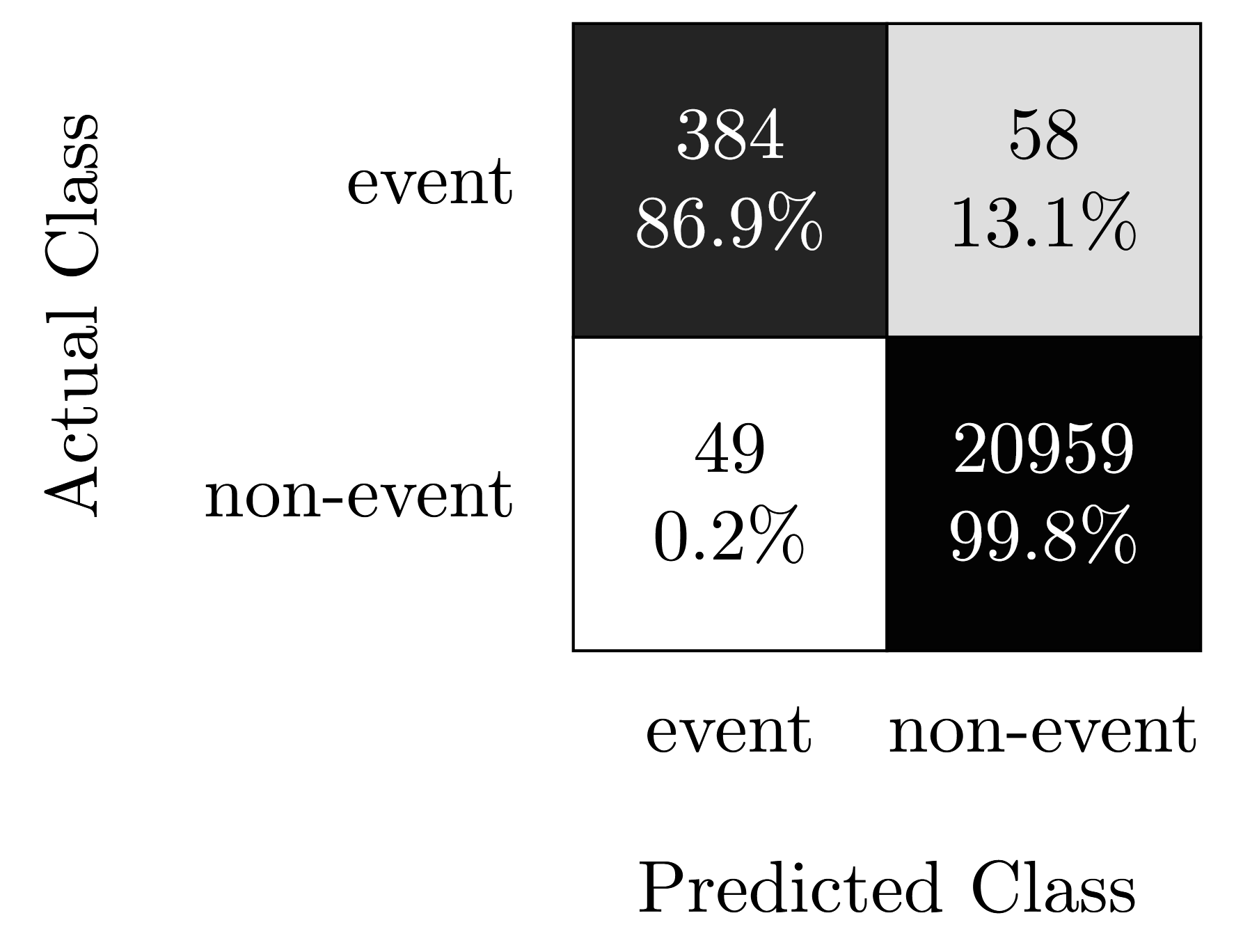}
    \caption{Confusion matrix for the L1C data.}
    \label{fig:L1CconMatrix}
    \end{subfigure}
    \caption{Confusion matrix for the raw (a) and L1C (b) datasets, showing the number of samples and the percentage for that class. The background colour shading represents the class percentage, with white for 0\% and black for 100\%.}
    \label{fig:conMatrices}
\end{figure}

\Cref{tab:results} also reports nearly identical recall performance for both the raw and L1C data at 0.867 and 0.869, respectively. The confusion matrices for the raw and L1C data are shown in \cref{fig:conMatrices}, where it can be seen that the difference in recall is due to only 1 more False Negative prediction in the raw data compared to the L1C. The real difference in performance between the raw and L1C datasets is due to the number of False Positives as highlighted by the difference in the precision values of 0.763 and 0.887, respectively. The number of False Positives is more than double in raw compared to L1C as seen in \cref{fig:conMatrices}, caused by the extra noise in the raw dataset.

\begin{figure}
    \centering
    \includegraphics[width=0.98\linewidth]{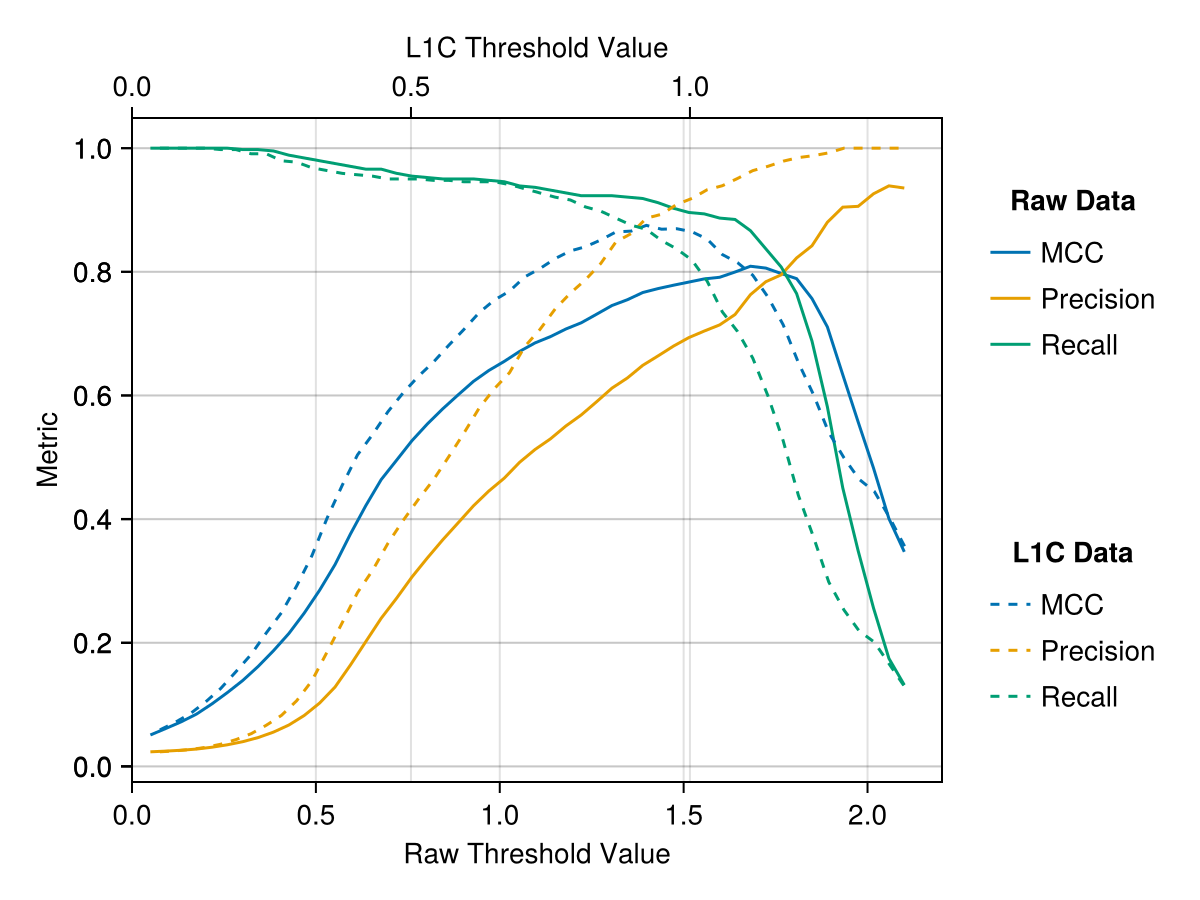}
    \caption{Effect of varying the threshold value on the different evaluation metrics; MCC (blue), precision (orange) and recall (green). Results for the raw data are shown with a solid line (x-axis values on bottom) and the L1C data results are shown with a dashed line (x-axis values on top).}
    \label{fig:thresholdCurve}
\end{figure}

Note, however, that the threshold value could be changed to increase either the precision or recall, depending on mission objectives and constraints when deployed. In this study, threshold values were chosen to maximise the MCC score. \Cref{fig:thresholdCurve} shows how the MCC, precision and recall values vary as the threshold value is adjusted for both the raw data (solid lines) and the L1C data (dashed lines). It can be seen that the area between the precision curves of the two datasets is larger than that of the recall curves, showing how the precision, and hence False Positive rate, is the main difference between the raw and L1C results.

\begin{figure*}
  \centering
  \begin{subfigure}{0.95\linewidth}
      \includegraphics[width=\linewidth]{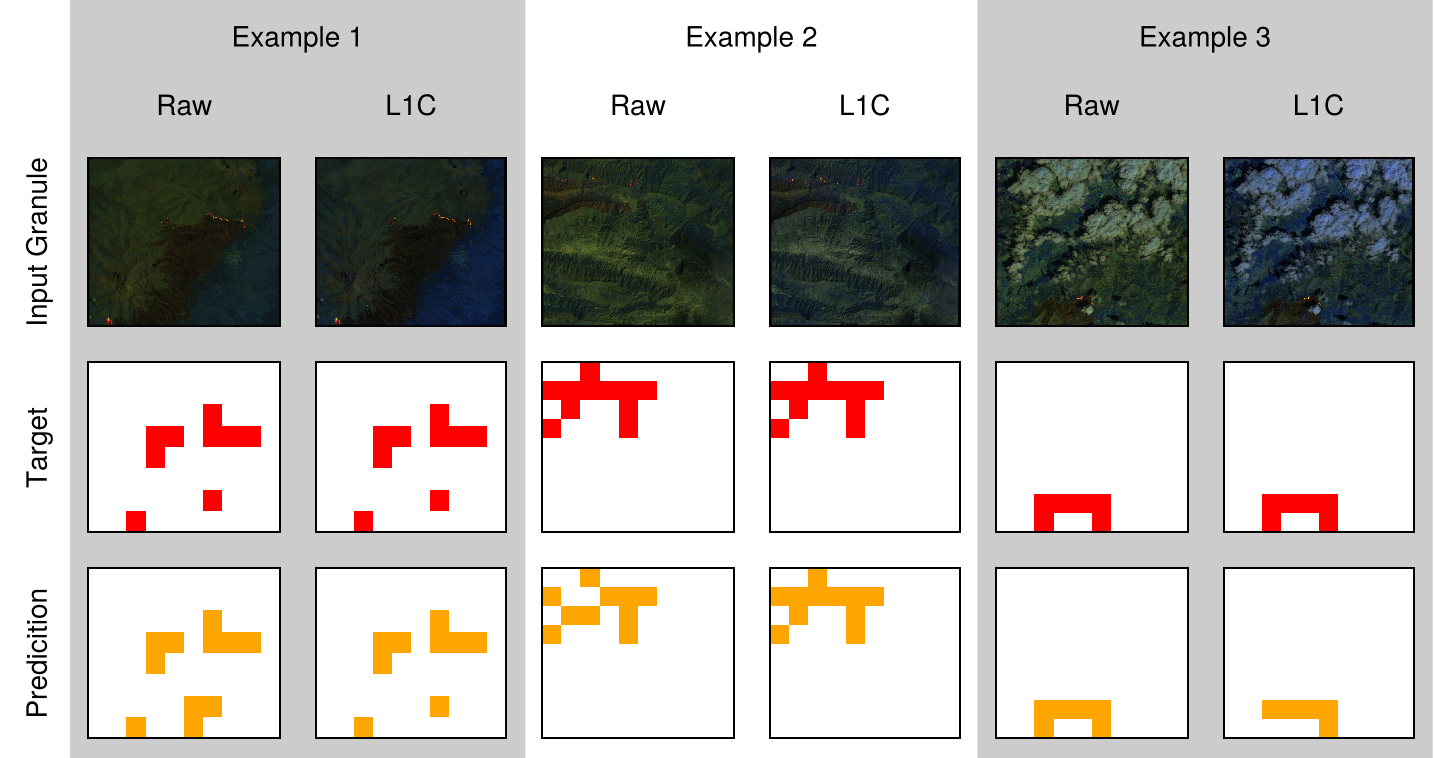}
      \caption{Three example fire thermal hotspots.}
  \end{subfigure}

  \vspace{2mm}
  \rule{\linewidth}{0.5pt}
  \vspace{3mm}
  
  \begin{subfigure}{0.95\linewidth}
      \includegraphics[width=\linewidth]{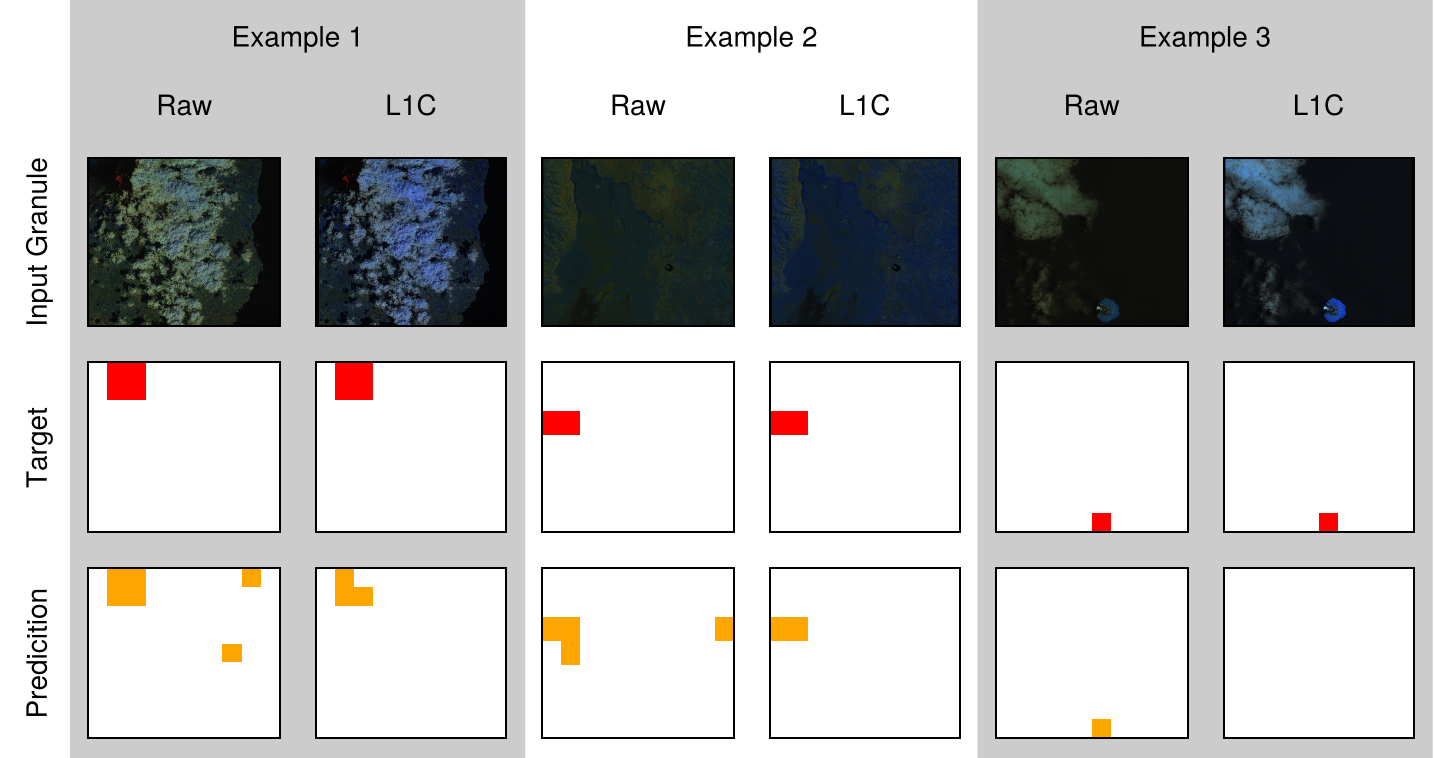}
      \caption{Three example volcano thermal hotspots.}
  \end{subfigure}
  \caption{Three examples for fires (a) and volcanoes (b) for both the raw and L1C datasets. Each example shows an RGB image, using bands B8A, B11 and B12 (top row), along with the target output (middle row), where tiles containing thermal anomalies are shown in red, and the predicted output (bottom row), where tiles predicted by the PANN to have thermal anomalies are shown in orange.}
  \label{fig:exampleResults}
\end{figure*}

\Cref{fig:exampleResults} shows 3 examples of fires and volcanoes each, for both the raw and L1C datasets. Each example, an RGB image using bands B8A, B11 and B12, a target mask and a predicted thermal mask, with tiles containing thermal anomalies highlighted in red and tiles predicted to have thermal anomalies highlighted in orange. The examples with significant differences between the raw and L1C predictions were chosen. 
It can be seen that the largest difference between raw and L1C is the increased number of False Positives as already noted.
Although the L1C data gives an overall higher performance, it is not necessarily the case that the L1C data always gives an equivalent or better result than the raw data; an example of such a case is shown in Example 3 for both the fires and volcanoes in \cref{fig:exampleResults}, where the model fails to predict thermal anomaly tiles when using the L1C data, but detects them when using the raw data. 

\subsection{Model Compute Resources}

\Cref{tab:computeResources} reports the time and maximum memory required to process a granule, for the steps shown in \cref{fig:modelSetup}, as well as a corresponding estimate for the hardware nanowire network device, which the PANN is based on. The most significant difference between the PANN and the estimated hardware is the time it takes to process a granule, with average times of $2.44\pm0.09$~\,s and $0.1290\pm0.0002$~\,s, respectively. Importantly, the time for the sensor to acquire a granule is 3.6~\,s, 
which means both the PANN model and hardware should be able to process a single granule within the acquisition time. Additionally, the hardware should be able  to operate in real time, as it could process all 12 granules within the acquisition time. While in this study the PANN model was run on hardware that would not be available for deployment in space, we used only 16 CPU cores to limit computing resources and there is the potential to deploy the PANN on space certified GPUs or custom accelerator chips to further speed up processing time. 
Likewise, the estimated hardware requirement of $0.393\pm0.004$~\,Gb is substantially less than the memory requirement of $0.673\pm0.007$~\,Gb for the PANN model. We note that the required memory includes loading all necessary programs and packages to run the model, as well as the memory size of the model itself, as this is a more accurate measure of the operational implementation. We also note that both the PANN model and estimated hardware used 32 bit precision and could be further reduced to decrease memory size. Regardless, the memory size of $0.673\pm0.007$~\,Gb for the PANN model is still within a reasonable size for onboard deployment on modern hardware architectures. 

A major advantage of the hardware nanowire network device is its low operational power and energy consumption. Using measured voltage and conductance values from a previous image classification study \cite{Zhu2023}, the average power is $\sim 25\,\mu$W. The estimated time to process a granule, 0.129~\,s, then implies an energy consumption of just $3.23\,\mu$J per granule. Furthermore, there is no additional energy overhead associated with training time.

\begin{table}
  \caption{Compute resources calculated for the PANN model and estimated for the hardware nanowire network device, including time to process a granule in seconds and the total memory requirement of the model and all required software and packages. For all values the $\pm$SEM is over 5 runs.}
  \label{tab:computeResources}
  \centering
  \begin{tabular}{@{}lcc@{}}
    \toprule
     & PANN model & Estimated Hardware \\
    \midrule
    Time (s / granule) & $2.44\pm0.09$ & $0.1290\pm0.0002$ \\
    Memory (GB) &  $0.673\pm0.007$ & $0.393\pm0.004$ \\
    \bottomrule
  \end{tabular}
\end{table}
\section{Conclusion}
\label{sec:conclusion}

This study applied a PANN model to detect thermal hotspot anomalies from EO images, as a proof of concept towards onboard processing. The model was evaluated on the THRawS dataset at two different processing levels, raw and L1C from the Sentinel-2 satellites. The results confirmed that the L1C data gave the best model performance with an MCC of 0.875, however, the raw data is comparable at 0.809. This demonstrates that the PANN model is capable of extracting meaningful features from the noisier raw data, which is essential for any onboard processing. 
Additionally, the time required for the PANN model to process two granules, one for each spectral band used in this study, was found to be $2.44\pm0.09$~\,s, using only CPUs. This is within the acquisition time of 3.6~\,s but does not include all 12 granules acquired by the sensor onboard the Sentinel-2 satellites. For the hardware nanowire network, on which the PANN is based, the projected processing time was $0.1290\pm0.0002$~\,s, well below the acquisition time and could process all 12 granules in real time. 
The memory requirements, for both the PANN and estimated hardware, were within a realistic size for onboard deployment at $0.673\pm0.007$~\,Gb and $0.393\pm0.004$~\,Gb, respectively, albeit on the high side. Further memory optimisation, such as quantising the model from 32-bit to 16-bit precision, could make the memory requirements more feasible. Altogether, these results show that the PANN model offers a promising approach towards low-latency and resource-efficient thermal event detection and is a viable candidate for processing EO data onboard satellites.

Additionally and most importantly, as demonstrated in this thermal hotspot detection task, the PANN model can be implemented in a way that does not require training. This allows the model to be used when labelled data is not available or is too expensive to label for the sensor being used. As such, unlike most ML models, the PANN does not need retraining with more new data due to domain shift or sensor drift and at most only requires the threshold value to be adjusted instead of the entire model needing to be retrained.

Future work includes implementing the PANN model on space certified hardware that would be deployed onboard satellites, particularly GPU architectures and custom accelerator chips to further speed up processing times. The implementation of the hardware nanowire network device should also be conducted to verify the projected time, memory and energy requirements calculated in this study. 
As the PANN model is training free, it is sensor agnostic and could generalise to different sensors or handle sensor drift. The model should therefore be applied to thermal hotspot tasks using data from different sensors to determine if this is the case. Additionally, the generalisability of the model to handle raw sensor data could be further investigated by applying the model to a variety of tasks beyond just thermal hotspot detection.

\section*{Acknowledgements}
This research was undertaken with the assistance of resources from the National Computational Infrastructure (NCI Australia), an NCRIS enabled capability supported by the Australian Government. S.S. is supported by an Australian Government Research Training Program (RTP) Scholarship.

{
    \small
    \bibliographystyle{ieeenat_fullname}
    \bibliography{main}
}

\end{document}